\documentclass[letterpaper]{article} 
\usepackage{aaai2026}  
\nocopyright
\usepackage{times}  
\usepackage{helvet}  
\usepackage{courier}  
\usepackage[hyphens]{url}  
\usepackage{graphicx} 
\usepackage{natbib}  
\usepackage{caption} 
\usepackage{algorithm}
\usepackage{algorithmic}

\usepackage{newfloat}
\usepackage{listings}
\DeclareCaptionStyle{ruled}{labelfont=normalfont,labelsep=colon,strut=off} 
\floatstyle{ruled}
\newfloat{listing}{tb}{lst}{}
\floatname{listing}{Listing}

\usepackage{booktabs}
\usepackage{pifont}
\usepackage{colortbl}
\usepackage{xcolor}
\usepackage{fontawesome5}
\newcommand{\yes}{\cellcolor{green!15}\textcolor{green!70!black}{\ding{51}}}
\newcommand{\no}{\cellcolor{red!15}\textcolor{red!70!black}{\ding{55}}}
\title{Who Gets Flagged? The Pluralistic Evaluation Gap in AI Content Watermarking}
\author {
    Alexander Nemecek,
    Osama Zafar,
    Yuqiao Xu,
    Wenbiao Li,
    Erman Ayday
}
\affiliations {
    Case Western Reserve University\\
    \{ajn98, oxz23, yxx914, wxl387, exa208\}@case.edu
}

\usepackage{bibentry}

\begin{document}

\maketitle

\begin{abstract}
Watermarking is becoming the default mechanism for AI content authentication, with governance policies and frameworks referencing it as infrastructure for content provenance. Yet across text, image, and audio modalities, watermark signal strength, detectability, and robustness depend on statistical properties of the content itself, properties that vary systematically across languages, cultural visual traditions, and demographic groups. We examine how this content dependence creates modality-specific pathways to bias. Reviewing the major watermarking benchmarks across modalities, we find that, with one exception, none report performance across languages, cultural content types, or population groups. To address this, we propose three concrete evaluation dimensions for pluralistic watermark benchmarking: cross-lingual detection parity, culturally diverse content coverage, and demographic disaggregation of detection metrics. We argue that watermarking is part of the pluralistic alignment pipeline and should be held to the same evaluation standards. We connect this to governance frameworks currently mandating watermarking deployment without requiring fairness evaluation. Our position is that evaluation must precede deployment, and that the same bias auditing requirements applied to AI models should extend to the verification layer.
\end{abstract}


\section{Introduction}
Watermarking is becoming the default mechanism for AI content authentication with governance acting accordingly. The European Union AI Act~\cite{eu_ai_act_2024}, China's Measures for Labeling AI-Generated Synthetic Content~\cite{china_ai_labeling_2025}, and, until its rescission in January 2025~\cite{eo14148_2025}, United States Executive Order 14110~\cite{eo14110_2023} have all referenced watermarking as infrastructure for content provenance. Major platforms including~\citet{google_deepmind_synthid},~\citet{meta_seal}, and~\citet{openai_ai_classifier_2023} have deployed watermarking or provenance tooling across their generative products. Yet as adoption accelerates on both regulatory and industry fronts, a question has gone largely unasked: who do these systems disproportionately affect?

The question matters because watermarking is not a neutral stamp applied uniformly to all content. Across modalities, watermark signal strength, detectability, and robustness depend on statistical properties of the content itself~\cite{piet2025markmywords, an2024waves, liu2024audiomarkbench}. These properties vary systematically across languages, cultural visual traditions, and linguistic communities~\cite{liu2024audiomarkbench, shira2025representation}. If watermarking becomes the primary mechanism by which AI-generated content is authenticated, then biased watermarking produces biased authentication. In agentic pipelines the disparity propagates: an agent that gates retrieval or moderation on provenance signals inherits the detector's bias.

Current evaluation practice is not equipped to surface this problem. The major watermark benchmarks (Table~\ref{tab:benchmarks}) focus on detectability, robustness to adversarial attack, output quality preservation, and computational overhead. All but one are evaluated under implicitly monocultural assumptions~\cite{agnew2024sound, mohamed2025multilingual}. Yet early evidence already demonstrates cause for concern: under identical conditions, watermark detection rates vary significantly across linguistic communities~\cite{xie2025watermark, xie2025silent}.

Pluralistic alignment seeks to align AI systems with the diversity of human values rather than a single averaged preference~\cite{sorensen2024roadmap}. \citeauthor{sorensen2024roadmap} formalize three operationalizations; the one our argument parallels is distributional pluralism, calibrating to the distribution of a target population. This program has been pursued almost exclusively at the level of model outputs (i.e., what the model says, whose preferences shape it, how disagreement is represented). We argue that pluralistic alignment is incomplete without extending the same scrutiny to the verification layer that determines whose outputs get authenticated.
 
The same governance frameworks effectively bind alignment requirements to watermarking: a generative system can pass a distributional pluralism evaluation while the watermarking layer that authenticates its outputs produces systematically different detection rates across the very populations the alignment effort was meant to serve. This is the verification-layer analog of \emph{pluralistic value-washing}~\cite{kasirzadeh2024plurality}, a superficially pluralistic pipeline whose binding component is monistic. If first-order alignment choices are made pluralistically but the second-order infrastructure that operationalizes them in deployment is evaluated only on aggregate, monocultural benchmarks, the apparent pluralism of the upstream system is silently flattened downstream.

Our three proposed evaluation dimensions: cross-lingual detection parity, culturally diverse content coverage, and demographic disaggregation, are the verification-layer counterpart to distributional pluralism for generation. They ask not whether a watermark works on average, but whether it works comparably across the populations a pluralistically-aligned system is designed to serve. We document existing disparities where empirical evidence exists, identify structural risk where evaluation is absent, and connect these to the governance frameworks that are mandating watermarking deployment without requiring fairness evaluation.

\section{How Watermarking Works}
Across modalities, watermarking embeds a statistical signal into AI-generated content during or after generation, and a detector later recovers that signal to verify provenance~\cite{zhao2025sok}. In every case, watermark strength and detectability depend on properties of the content itself. The dependence that makes watermarking effective also makes it vulnerable to systematic bias when content properties vary across languages, cultures, and communities.

In \textbf{text}, the dominant approach modifies token sampling during generation~\cite{kirchenbauer2023watermark, dathathri2024scalable, nemecek2024topic}, with post-hoc alternatives operating through synonym substitution~\cite{chang2024postmark} or syntactic restructuring~\cite{rizzo2016homoglyph}. Detection of the watermark across approaches depends on the entropy of the token distribution. The signal weakens in low-entropy contexts where few plausible next tokens leave room to bias selection without degrading quality~\cite{piet2025markmywords, wu2025analyzing}. In \textbf{images}, latent-space methods embed the signal during diffusion sampling~\cite{wen2023tree, yang2024gaussian} while post-hoc methods embed a payload into pixel data~\cite{gowal2025synthid, tancik2020stegastamp, bui2025trustmark}. Both methods encode the watermark in frequency-domain representations optimized for imperceptibility and robustness to compression, so high-frequency detail, color distribution, texture complexity, and spatial structure all shape embedding and detection~\cite{qiu2024evaluating}. \textbf{Audio} systems such as AudioSeal~\cite{roman2024proactive} likewise operate in the frequency domain, embedding imperceptible waveform signals that must survive noise injection, speed modification, and re-encoding~\cite{chen2023wavmark}, so the acoustic properties of speech (i.e., pitch, spectral energy distribution, prosodic contour) shape how the watermark is embedded and recovered~\cite{ozer2025comprehensive}.

\begin{table*}[t]
\centering
\small
\begin{tabular}{llccc}
\toprule
\textbf{Benchmark} & \textbf{Modality} & \textbf{Multilingual/Cross-cultural} & \textbf{Demographic Reporting} & \textbf{Bias Discussed} \\
\midrule
MarkMyWords~\cite{piet2025markmywords} & \faFont\ Text & \no & \no & \no \\
WaterBench~\cite{tu2024waterbench} & \faFont\ Text & \no & \no & \no \\
WaterPark~\cite{liang2024watermark} & \faFont\ Text & \no & \no & \no \\
WAVES~\cite{an2024waves} & \faImage\ Image & \no & \no & \no \\
W-Bench~\cite{lu2024robust} & \faImage\ Image & \no & \no & \no \\
VideoMarkBench~\cite{jiang2025videomarkbench} & \faVideo\ Video & \no & \no & \no \\
AudioMarkBench~\cite{liu2024audiomarkbench} & \faVolumeUp\ Audio & \yes & \yes & \yes \\
\bottomrule
\end{tabular}
\caption{Pluralistic evaluation of watermarking benchmarks. Multilingual/Cross-cultural indicates whether evaluation includes non-English languages or non-Western content. Demographic Reporting indicates whether detection metrics are disaggregated by population group.}
\label{tab:benchmarks}
\end{table*}

\section{Pluralistic Failures Across Modalities}\label{bias-gaps}
The evaluation landscape for watermarking reveals a consistent pattern in which pluralistic evaluation is treated as out of scope. Table~\ref{tab:benchmarks} summarizes the pluralistic evaluation dimensions of major watermarking benchmarks across text, image, audio, and video. With a single exception, none report performance across languages, cultural content types, or demographic groups. The table deliberately covers benchmark suites rather than individual studies: we count as a benchmark a general-purpose evaluation suite that tests multiple watermarking methods under a common protocol of perturbations and metrics and is intended as a reusable standard, which is what makes such suites the field's default reporting practice. The list is representative rather than exhaustive, covering the principal general-purpose suites in each modality. Targeted analyses~\cite{he2024can, al-ghanim-etal-2025-evaluating, mohamed2025multilingual, xie2025watermark} already show that these dimensions are measurable; their measurements have simply not been absorbed into the benchmarks by which watermarking systems are validated. The evidence that follows is accordingly tiered: documented disparate impact in text, structurally likely but untested bias in images, and in audio a single benchmark that found disparities where no other has looked.

\textbf{Text.} Bias in text watermarking is not hypothetical.~\citet{xie2025silent} demonstrates that under identical permitted AI assistance guidelines, essays by non-native English speakers carry stronger watermark signals than those by native speakers, resulting in higher false positive rates at standard detection thresholds. AI systems make more extensive modifications to non-native text, producing more watermarked tokens at more high-entropy positions; the bias is structural, arising from the interaction between the writer's linguistic background and the embedding mechanism rather than from the detector.~\citet{xie2025watermark} formalize this disparity, showing that non-native speakers' essays exhibit lower watermark p-values even when both groups use only permitted grammatical editing, and propose conformal methods to calibrate per-group detection thresholds.

Beyond the generation process, tokenization introduces a second axis of disparity. Language model tokenizers trained on English-heavy corpora produce efficient single-token representations for common English words while fragmenting equivalent content in non-Latin scripts into multiple subword units~\cite{petrov2023language, ahia-etal-2023-languages}. Since watermark detection aggregates signal across tokens, the same semantic content expressed in Chinese or Arabic may produce different detection statistics than its English equivalent.~\citet{he2024can} confirm a related failure: token-distribution watermarks lose detectability when text is translated across languages, with detection falling to near chance. Despite this growing body of evidence, the major text watermarking benchmarks (Table~\ref{tab:benchmarks}) evaluate exclusively on English corpora and report no demographic disaggregation of detection rates.

\textbf{Images.} 
To our knowledge, no published work has evaluated whether image watermark detection performance varies as a function of cultural content, subject demographics, or non-Western visual traditions. The gap exists despite clear reason to investigate, as Google's SynthID-Image~\cite{gowal2025synthid} acknowledges that learned perceptual quality metrics are biased toward the content distributions they were developed on, in practice ImageNet-style natural images. If the metrics used to assess watermark imperceptibility cannot reliably evaluate culturally diverse content, then differential watermark behavior across visual traditions would be invisible to current evaluation. 

The primary image watermark robustness benchmark, WAVES~\cite{an2024waves}, tests perturbations reflecting Western platform conventions (e.g., JPEG compression, cropping, rotation). These operations are not demographically neutral. Lossy JPEG compression alone has been shown to disproportionately degrade face recognition performance on darker skin tones (up to 34\% accuracy reduction) via chroma subsampling in the frequency domain~\cite{yucer2022does}. This is indirect evidence, from recognition rather than watermark detection, but it shows that the primitives watermark robustness tests rely on already encode demographic asymmetries before any watermark is introduced. Absent from evaluation are compression codecs and processing pipelines common on non-Western platforms such as WeChat and LINE~\cite{li2020anti}, or visual content types that fall outside ImageNet-derived distributions, including calligraphic scripts, intricate textile patterns, and non-photographic art traditions. The problem compounds at the generative layer: text-to-image systems default to Western-centric depictions under culturally neutral prompts, generating predominantly white male subjects and misrepresenting non-Western cultures~\cite{ghosh-caliskan-2023-person, bianchi2023easily}. If the generated content is culturally skewed and the watermark is trained and evaluated on that same skewed distribution, the verification layer inherits the model's biases while introducing potential new ones through the embedding process itself. We cannot claim image watermarking is fair because we have not evaluated it for fairness. The mechanisms predict differential behavior, and the evaluation infrastructure was not designed to detect it.

\textbf{Audio.} Audio watermarking presents a unique case of one benchmark that tested for demographic disparities and found them. AudioMarkBench~\cite{liu2024audiomarkbench} evaluated watermark robustness across 25 languages and stratified results by biological sex and age group. The findings showed that under certain perturbations, female speech exhibited higher false positive rates for watermark forgery than male speech. Language-level variation was also detected, with languages such as Georgian and Esperanto showing anomalously different false negative rates depending on the watermarking method and perturbation type. Yet AudioMarkBench remains the sole exception in a field that otherwise evaluates exclusively on English-language generated audio. The structural reasons to expect broader disparities are clear. Languages such as Mandarin, Vietnamese, and Yoruba encode semantic meaning in pitch contours~\cite{kaur2021automatic}, presenting acoustic profiles fundamentally different from English prosody. Frequency-domain watermarks that modify or depend on spectral characteristics~\cite{wen2025sok} could interact with tonal information in ways that current benchmarks do not capture. The gap extends to deployment infrastructure as well: audio sharing platforms, compression formats, and processing pipelines vary globally, yet robustness evaluation is limited to Western codecs and transformations. Audio watermarking is already deployed in production and referenced in governance mandates~\cite{fernandez2025audioseal, eu_ai_act_2024}, so populations whose languages and acoustic environments differ from the evaluation distribution are subject to a verification system whose behavior on their content is unknown.

\section{Toward Pluralistic Watermark Evaluation}\label{toward}
The gaps documented above are not inevitable but choices that can be revised, as AudioMarkBench demonstrates. The dimensions that follow are, we argue, minimum requirements for any watermark benchmark deployed under governance mandates. Throughout, detection parity means parity in each error rate separately rather than in aggregate accuracy. For example, the over-flagging of non-native writers and the forgery false positives on female speech documented above are both false positive disparities on unwatermarked content, while detection lost to translation is a false negative disparity on watermarked content, and a benchmark can be at parity on one while failing the other.

\textbf{Cross-Lingual Detection Parity.} Watermark benchmarks must evaluate detection performance across languages and script families as a core reporting requirement. For text, this means testing across languages with different tokenization efficiencies and morphological structures~\cite{ahia-etal-2023-languages} and reporting per-language error rates~\cite{al-ghanim-etal-2025-evaluating}. Audio evaluation must extend beyond English speech to include tonal and typologically diverse languages. For images, text-bearing content such as signage and calligraphic elements varies across scripts in ways that may plausibly interact with frequency-domain embedding.

\textbf{Culturally Diverse Content.} Benchmark content must extend beyond the distributions on which watermarking systems are developed and tested. For images, this means including visual traditions outside ImageNet-derived distributions~\cite{de2019does, NEURIPS2022_5474d9d4} and content generated under culturally specific prompts. For audio, benchmarks must include speech from naturalistic acoustic environments, capturing the reverberation, background noise, and equipment common across regions~\cite{barker2018fifth}. For text, evaluation corpora should include content across genres and rhetorical conventions that vary culturally, not only formal English prose~\cite{bender2021dangers}. The goal is not exhaustive coverage of every tradition but representative sampling sufficient to surface systematic disparities before deployment.

\textbf{Demographic Disaggregation.} Watermark detection metrics must be reported stratified by population rather than as single aggregates, following established practice in AI fairness evaluation~\cite{pmlr-v81-buolamwini18a} that watermarking has not adopted. For text, this means disaggregating by the linguistic background of the writer, given the native/non-native disparities documented above~\cite{xie2025watermark}. For audio, AudioMarkBench's stratification by biological sex and age already provides the template. For images, disaggregating by the demographic composition of generated subjects and the cultural origin of visual content would establish whether embedding and detection behave uniformly across depicted populations.

\textbf{Minimum Reporting.} Concretely, a benchmark meeting these dimensions reports, per language and population group: false positive and false negative rates separately at fixed thresholds, with confidence intervals and group sample sizes, since the two errors fall on different parties (the writer whose work is misattributed versus the provenance guarantee that synthetic content defeats); each under clean conditions and after standard perturbations, including translation for text, since detection lost to perturbation is a distinct failure from baseline miscalibration; and over a language set chosen to span script families and tokenization efficiencies rather than by convenience, with demographic and cultural strata matched to those already used in audits of the generative layer~\cite{hall2024towards} and group gaps reported as differences or ratios in per-group error rates.

\textbf{Robustness Trade-offs.} The dimensions above interact with the robustness and imperceptibility goals that watermark benchmarks have prioritized. Per-group threshold calibration can equalize false positive rates but requires group information at detection time, raising privacy concerns about classifying users by demographic background~\cite{xie2025watermark}. Adaptive embedding strength to equalize detectability across content types may compromise imperceptibility on content or robustness on others. Cross-lingual consistency requires watermarks robust to translation, which standard token-distribution methods do not provide~\cite{he2024can}; semantic-invariant variants offer partial recovery.

These trade-offs cannot be characterized without group-level measurement. What aggregate benchmarks report as robustness is, in practice, robustness for the dominant distribution; a system that loses some aggregate detection but gains detectability for underserved groups is not strictly worse, only differently calibrated. Which fairness criterion is appropriate (equal false positive rates, equalized odds, proportional impact thresholds) will vary by deployment and regulatory jurisdiction, but no criterion can be applied, and no trade-off frontier mapped, until stratified performance is reported, which the field does not yet do.

\section{Policy Implications}

The evaluation gaps are not purely academic concerns. The European Union AI Act~\cite{eu_ai_act_2024} and China's synthetic-content labeling measures~\cite{china_ai_labeling_2025} both mandate some form of content marking or traceability for AI-generated outputs; United States Executive Order 14110~\cite{eo14110_2023} did the same until its rescission in January 2025~\cite{eo14148_2025}. None of the three require that detection performance be evaluated across languages, cultural content types, or demographic groups. In each case, watermarking is treated as neutral infrastructure rather than as a system whose behavior varies with the content it processes. The asymmetry is stark: the EU AI Act requires bias evaluation for general-purpose and high-risk AI systems, yet exempts watermarking from equivalent scrutiny~\cite{nemecek2025watermarking}.

Closing this gap does not require novel regulatory architecture, only extending existing audit requirements to cover watermarking~\cite{benerofe2025ai}. This means three additions: first, including watermark detection parity in the conformity assessments already required for high-risk AI systems; second, requiring per-language and per-demographic reporting of false positive and negative rates as a condition of deployment; and third, mandating that watermark robustness evaluation include compression codecs and platform pipelines representative of global deployment contexts. All three mirror the disaggregated reporting and demographic auditing already required of other AI system components~\cite{nist_ai_rmf_2023}; the only change is holding watermarking to the same standard.

\section{Conclusion}

This paper has argued that watermarking, increasingly mandated as the verification layer for AI-generated content, carries the same potential for disparate impact as the generative systems it is meant to authenticate. The evidence, direct for text and audio and mechanistic for images, indicates that content-dependent embedding produces content-dependent outcomes, and content varies systematically across languages, cultures, and communities. The field has optimized watermarking for robustness and imperceptibility while treating the populations it affects as uniform. They are not. The evaluation dimensions we propose are not aspirational, as a single benchmark has already demonstrated their feasibility. What is missing is not capability but priority. Pluralistic alignment cannot end at generation; the verification layer that authenticates whose content counts is part of the pipeline, and must be evaluated accordingly.

\section{Acknowledgments}
This work was supported by the National Science Foundation (NSF) under grant numbers OAC-2112606, 2427505, and 2141622, and by the National Library of Medicine of the National Institutes of Health (NIH) under award number R01LM014520.

\bibliography{aaai2026}

@misc{eu_ai_act_2024,
  title        = {Regulation ({EU}) 2024/1689 of the {European Parliament} and of the {Council} of 13 June 2024 Laying Down Harmonised Rules on Artificial Intelligence and Amending Regulations ({EC}) No 300/2008, ({EU}) No 167/2013, ({EU}) No 168/2013, ({EU}) 2018/858, ({EU}) 2018/1139 and ({EU}) 2019/2144 and Directives 2014/90/{EU}, ({EU}) 2016/797 and ({EU}) 2020/1828 (Artificial Intelligence Act)},
  author       = {{European Parliament and Council of the European Union}},
  journal      = {Official Journal of the European Union},
  volume       = {L},
  year         = {2024},
  url          = {https://eur-lex.europa.eu/eli/reg/2024/1689/oj/eng},
}

@misc{eo14110_2023,
  title        = {Executive Order 14110: Safe, Secure, and Trustworthy Development and Use of Artificial Intelligence},
  author       = {{Executive Office of the President}},
  year         = {2023},
  note         = {88 FR 75191, Document 2023-24283},
  url          = {https://www.federalregister.gov/documents/2023/11/01/2023-24283/safe-secure-and-trustworthy-development-and-use-of-artificial-intelligence}
}

@misc{eo14148_2025,
  title        = {Executive Order 14148: Initial Rescissions of Harmful Executive Orders and Actions},
  author       = {{Executive Office of the President}},
  year         = {2025},
  note         = {90 FR 8237, Document 2025-01901},
  url          = {https://www.federalregister.gov/documents/2025/01/28/2025-01901/initial-rescissions-of-harmful-executive-orders-and-actions}
}

@misc{china_ai_labeling_2025,
  title        = {Measures for Labeling of AI-Generated Synthetic Content},
  author = {{Cyberspace Administration of China} and {Ministry of Industry and Information Technology} and {Ministry of Public Security} and {State Administration of Radio and Television}},
  year         = {2025},
  note         = {Document No. State Information Office Tongzi [2025] No. 2},
  url          = {https://www.chinalawtranslate.com/en/ai-labeling/}
}

@article{dathathri2024scalable,
  title={Scalable watermarking for identifying large language model outputs},
  author={Dathathri, Sumanth and See, Abigail and Ghaisas, Sumedh and Huang, Po-Sen and McAdam, Rob and Welbl, Johannes and Bachani, Vandana and Kaskasoli, Alex and Stanforth, Robert and Matejovicova, Tatiana and others},
  journal={Nature},
  volume={634},
  number={8035},
  pages={818--823},
  year={2024},
  publisher={Nature Publishing Group UK London}
}

@misc{google_deepmind_synthid,
  title        = {{SynthID}},
  author       = {{Google DeepMind}},
  year         = {2026},
  url          = {https://deepmind.google/models/synthid/},
  urldate      = {2026-04-06}
}

@misc{meta_seal,
  title        = {{Meta Seal}: State-of-the-Art, Open Source Invisible Watermarking},
  author       = {{Meta FAIR}},
  year         = {2026},
  url          = {https://facebookresearch.github.io/meta-seal/},
  urldate      = {2026-04-06}
}

@misc{openai_ai_classifier_2023,
  title        = {New {AI} Classifier for Indicating {AI}-Written Text},
  author       = {{OpenAI}},
  year         = {2023},
  month        = jan,
  day          = {31},
  url          = {https://openai.com/index/new-ai-classifier-for-indicating-ai-written-text/},
  urldate      = {2026-04-06},
  note         = {Discontinued July 20, 2023 due to low accuracy}
}

@article{liu2024audiomarkbench,
  title={Audiomarkbench: Benchmarking robustness of audio watermarking},
  author={Liu, Hongbin and Guo, Moyang and Jiang, Zhengyuan and Wang, Lun and Gong, Neil Z},
  journal={Advances in Neural Information Processing Systems},
  volume={37},
  pages={52241--52265},
  year={2024}
}

@inproceedings{shira2025representation,
author = {Michel, Shira and Kaur, Sufi and Gillespie, Sarah Elizabeth and Gleason, Jeffrey and Wilson, Christo and Ghosh, Avijit},
title = {“It’s not a representation of me”: Examining Accent Bias and Digital Exclusion in Synthetic AI Voice Services},
year = {2025},
isbn = {9798400714825},
publisher = {Association for Computing Machinery},
address = {New York, NY, USA},
url = {https://doi.org/10.1145/3715275.3732018},
doi = {10.1145/3715275.3732018},
booktitle = {Proceedings of the 2025 ACM Conference on Fairness, Accountability, and Transparency},
pages = {228–245},
numpages = {18},
location = {
},
series = {FAccT '25}
}

@inproceedings{piet2025markmywords,
  title={Markmywords: Analyzing and evaluating language model watermarks},
  author={Piet, Julien and Sitawarin, Chawin and Fang, Vivian and Mu, Norman and Wagner, David},
  booktitle={2025 IEEE Conference on Secure and Trustworthy Machine Learning (SaTML)},
  pages={68--91},
  year={2025},
  organization={IEEE}
}

@inproceedings{tu2024waterbench,
  title={Waterbench: Towards holistic evaluation of watermarks for large language models},
  author={Tu, Shangqing and Sun, Yuliang and Bai, Yushi and Yu, Jifan and Hou, Lei and Li, Juanzi},
  booktitle={Proceedings of the 62nd Annual Meeting of the Association for Computational Linguistics (Volume 1: Long Papers)},
  pages={1517--1542},
  year={2024}
}

@article{liang2024watermark,
  title={Watermark under fire: A robustness evaluation of llm watermarking},
  author={Liang, Jiacheng and Wang, Zian and Hong, Spencer and Ji, Shouling and Wang, Ting},
  journal={arXiv preprint arXiv:2411.13425},
  year={2024}
}

@article{an2024waves,
  title={Waves: Benchmarking the robustness of image watermarks},
  author={An, Bang and Ding, Mucong and Rabbani, Tahseen and Agrawal, Aakriti and Xu, Yuancheng and Deng, Chenghao and Zhu, Sicheng and Mohamed, Abdirisak and Wen, Yuxin and Goldstein, Tom and others},
  journal={arXiv preprint arXiv:2401.08573},
  year={2024}
}

@article{jiang2025videomarkbench,
  title={VideoMarkBench: Benchmarking Robustness of Video Watermarking},
  author={Jiang, Zhengyuan and Guo, Moyang and Li, Kecen and Hu, Yuepeng and Wang, Yupu and Huang, Zhicong and Hong, Cheng and Gong, Neil Zhenqiang},
  journal={arXiv preprint arXiv:2505.21620},
  year={2025}
}

@article{lu2024robust,
  title={Robust watermarking using generative priors against image editing: From benchmarking to advances},
  author={Lu, Shilin and Zhou, Zihan and Lu, Jiayou and Zhu, Yuanzhi and Kong, Adams Wai-Kin},
  journal={arXiv preprint arXiv:2410.18775},
  year={2024}
}

@article{xie2025watermark,
  title={Watermark in the Classroom: A Conformal Framework for Adaptive AI Usage Detection},
  author={Xie, Yangxinyu and Chen, Xuyang and Ren, Zhimei and Su, Weijie J},
  journal={arXiv preprint arXiv:2507.23113},
  year={2025}
}

@misc{xie2025silent,
  title        = {Silent Discrimination: How {AI} Watermarking Systems Create Digital Accents in Non-Native {English} Writing},
  author       = {Yangxinyu Xie},
  year         = {2025},
  url          = {https://mackinstitute.wharton.upenn.edu/2025/silent-discrimination-how-ai-watermarking-systems-create-digital-accents-in-non-native-english-writing/},
  urldate      = {2026-04-06},
}

@inproceedings{hall2024towards,
  title={Towards geographic inclusion in the evaluation of text-to-image models},
  author={Hall, Melissa and Bell, Samuel J and Ross, Candace and Williams, Adina and Drozdzal, Michal and Soriano, Adriana Romero},
  booktitle={Proceedings of the 2024 ACM Conference on Fairness, Accountability, and Transparency},
  pages={585--601},
  year={2024}
}

@article{agnew2024sound,
  title={Sound check: Auditing audio datasets},
  author={Agnew, William and Barnett, Julia and Chu, Annie and Hong, Rachel and Feffer, Michael and Netzorg, Robin and Jiang, Harry H and Awumey, Ezra and Das, Sauvik},
  journal={arXiv preprint arXiv:2410.13114},
  year={2024}
}

@article{mohamed2025multilingual,
  title={Is Multilingual LLM Watermarking Truly Multilingual? A Simple Back-Translation Solution},
  author={Mohamed, Asim and Gubri, Martin},
  journal={arXiv preprint arXiv:2510.18019},
  year={2025}
}

@inproceedings{zhao2025sok,
  title={Sok: Watermarking for ai-generated content},
  author={Zhao, Xuandong and Gunn, Sam and Christ, Miranda and Fairoze, Jaiden and Fabrega, Andres and Carlini, Nicholas and Garg, Sanjam and Hong, Sanghyun and Nasr, Milad and Tramer, Florian and others},
  booktitle={2025 IEEE Symposium on Security and Privacy (SP)},
  pages={2621--2639},
  year={2025},
  organization={IEEE}
}

@inproceedings{kirchenbauer2023watermark,
  title={A watermark for large language models},
  author={Kirchenbauer, John and Geiping, Jonas and Wen, Yuxin and Katz, Jonathan and Miers, Ian and Goldstein, Tom},
  booktitle={International conference on machine learning},
  pages={17061--17084},
  year={2023},
  organization={PMLR}
}

@article{nemecek2024topic,
  title={Topic-based watermarks for large language models},
  author={Nemecek, Alexander and Jiang, Yuzhou and Ayday, Erman},
  journal={arXiv preprint arXiv:2404.02138},
  year={2024}
}

@inproceedings{chang2024postmark,
  title={Postmark: A robust blackbox watermark for large language models},
  author={Chang, Yapei and Krishna, Kalpesh and Houmansadr, Amir and Wieting, John Frederick and Iyyer, Mohit},
  booktitle={Proceedings of the 2024 Conference on Empirical Methods in Natural Language Processing},
  pages={8969--8987},
  year={2024}
}

@inproceedings{rizzo2016homoglyph,
author = {Rizzo, Stefano Giovanni and Bertini, Flavio and Montesi, Danilo},
title = {Content-preserving Text Watermarking through Unicode Homoglyph Substitution},
year = {2016},
isbn = {9781450341189},
publisher = {Association for Computing Machinery},
address = {New York, NY, USA},
url = {https://doi.org/10.1145/2938503.2938510},
doi = {10.1145/2938503.2938510},
booktitle = {Proceedings of the 20th International Database Engineering \& Applications Symposium},
pages = {97–104},
numpages = {8},
location = {Montreal, QC, Canada},
series = {IDEAS '16}
}

@article{wu2025analyzing,
  title={Analyzing and evaluating unbiased language model watermark},
  author={Wu, Yihan and Cui, Xuehao and Chen, Ruibo and Huang, Heng},
  journal={arXiv preprint arXiv:2509.24048},
  year={2025}
}

@article{wen2023tree,
  title={Tree-ring watermarks: Fingerprints for diffusion images that are invisible and robust},
  author={Wen, Yuxin and Kirchenbauer, John and Geiping, Jonas and Goldstein, Tom},
  journal={arXiv preprint arXiv:2305.20030},
  year={2023}
}

@inproceedings{bui2025trustmark,
  title={Trustmark: Robust watermarking and watermark removal for arbitrary resolution images},
  author={Bui, Tu and Agarwal, Shruti and Collomosse, John},
  booktitle={Proceedings of the IEEE/CVF International Conference on Computer Vision},
  pages={18629--18639},
  year={2025}
}

@inproceedings{yang2024gaussian,
  title={Gaussian shading: Provable performance-lossless image watermarking for diffusion models},
  author={Yang, Zijin and Zeng, Kai and Chen, Kejiang and Fang, Han and Zhang, Weiming and Yu, Nenghai},
  booktitle={Proceedings of the IEEE/CVF Conference on Computer Vision and Pattern Recognition},
  pages={12162--12171},
  year={2024}
}

@article{gowal2025synthid,
  title={SynthID-Image: Image watermarking at internet scale},
  author={Gowal, Sven and Bunel, Rudy and Stimberg, Florian and Stutz, David and Ortiz-Jimenez, Guillermo and Kouridi, Christina and Vecerik, Mel and Hayes, Jamie and Rebuffi, Sylvestre-Alvise and Bernard, Paul and others},
  journal={arXiv preprint arXiv:2510.09263},
  year={2025}
}

@inproceedings{tancik2020stegastamp,
  title={Stegastamp: Invisible hyperlinks in physical photographs},
  author={Tancik, Matthew and Mildenhall, Ben and Ng, Ren},
  booktitle={Proceedings of the IEEE/CVF conference on computer vision and pattern recognition},
  pages={2117--2126},
  year={2020}
}

@article{qiu2024evaluating,
  title={Evaluating Durability: Benchmark Insights into Multimodal Watermarking},
  author={Qiu, Jielin and Han, William and Zhao, Xuandong and Long, Shangbang and Faloutsos, Christos and Li, Lei},
  journal={arXiv preprint arXiv:2406.03728},
  year={2024}
}

@article{roman2024proactive,
  title={Proactive detection of voice cloning with localized watermarking},
  author={Roman, Robin San and Fernandez, Pierre and D{\'e}fossez, Alexandre and Furon, Teddy and Tran, Tuan and Elsahar, Hady},
  journal={arXiv preprint arXiv:2401.17264},
  year={2024}
}

@article{chen2023wavmark,
  title={Wavmark: Watermarking for audio generation},
  author={Chen, Guangyu and Wu, Yu and Liu, Shujie and Liu, Tao and Du, Xiaoyong and Wei, Furu},
  journal={arXiv preprint arXiv:2308.12770},
  year={2023}
}

@article{ozer2025comprehensive,
  title={A comprehensive real-world assessment of audio watermarking algorithms: Will they survive neural codecs?},
  author={{\"O}zer, Yigitcan and Choi, Woosung and Serr{\`a}, Joan and Singh, Mayank Kumar and Liao, Wei-Hsiang and Mitsufuji, Yuki},
  journal={arXiv preprint arXiv:2505.19663},
  year={2025}
}

@inproceedings{he2024can,
  title={Can watermarks survive translation? on the cross-lingual consistency of text watermark for large language models},
  author={He, Zhiwei and Zhou, Binglin and Hao, Hongkun and Liu, Aiwei and Wang, Xing and Tu, Zhaopeng and Zhang, Zhuosheng and Wang, Rui},
  booktitle={Proceedings of the 62nd Annual Meeting of the Association for Computational Linguistics (Volume 1: Long Papers)},
  pages={4115--4129},
  year={2024}
}

@article{petrov2023language,
  title={Language model tokenizers introduce unfairness between languages},
  author={Petrov, Aleksandar and La Malfa, Emanuele and Torr, Philip and Bibi, Adel},
  journal={Advances in neural information processing systems},
  volume={36},
  pages={36963--36990},
  year={2023}
}

@inproceedings{ahia-etal-2023-languages,
    title = "Do All Languages Cost the Same? Tokenization in the Era of Commercial Language Models",
    author = "Ahia, Orevaoghene  and
      Kumar, Sachin  and
      Gonen, Hila  and
      Kasai, Jungo  and
      Mortensen, David  and
      Smith, Noah  and
      Tsvetkov, Yulia",
    editor = "Bouamor, Houda  and
      Pino, Juan  and
      Bali, Kalika",
    booktitle = "Proceedings of the 2023 Conference on Empirical Methods in Natural Language Processing",
    month = dec,
    year = "2023",
    address = "Singapore",
    publisher = "Association for Computational Linguistics",
    url = "https://aclanthology.org/2023.emnlp-main.614/",
    doi = "10.18653/v1/2023.emnlp-main.614",
    pages = "9904--9923"
}

@inproceedings{bianchi2023easily,
  title={Easily accessible text-to-image generation amplifies demographic stereotypes at large scale},
  author={Bianchi, Federico and Kalluri, Pratyusha and Durmus, Esin and Ladhak, Faisal and Cheng, Myra and Nozza, Debora and Hashimoto, Tatsunori and Jurafsky, Dan and Zou, James and Caliskan, Aylin},
  booktitle={Proceedings of the 2023 ACM conference on fairness, accountability, and transparency},
  pages={1493--1504},
  year={2023}
}

@inproceedings{ghosh-caliskan-2023-person,
    title = "`Person' == Light-skinned, Western Man, and Sexualization of Women of Color: Stereotypes in Stable Diffusion",
    author = "Ghosh, Sourojit  and
      Caliskan, Aylin",
    editor = "Bouamor, Houda  and
      Pino, Juan  and
      Bali, Kalika",
    booktitle = "Findings of the Association for Computational Linguistics: EMNLP 2023",
    month = dec,
    year = "2023",
    address = "Singapore",
    publisher = "Association for Computational Linguistics",
    url = "https://aclanthology.org/2023.findings-emnlp.465/",
    doi = "10.18653/v1/2023.findings-emnlp.465",
    pages = "6971--6985"
}

@article{kaur2021automatic,
  title={Automatic Speech Recognition System for Tonal Languages: State-of-the-Art Survey: J. Kaur et al.},
  author={Kaur, Jaspreet and Singh, Amitoj and Kadyan, Virender},
  journal={Archives of Computational Methods in Engineering},
  volume={28},
  number={3},
  pages={1039--1068},
  year={2021},
  publisher={Springer}
}

@misc{fernandez2025audioseal,
  author       = {Pierre Fernandez},
  title        = {Proactive Detection of Voice Cloning with Localized Watermarking},
  howpublished = {\url{https://pierrefdz.github.io/publications/audioseal/}},
  year         = {2025},
  note         = {Accessed: 2026-01-28}
}

@article{nemecek2025watermarking,
  title={Watermarking Without Standards Is Not AI Governance},
  author={Nemecek, Alexander and Jiang, Yuzhou and Ayday, Erman},
  journal={arXiv preprint arXiv:2505.23814},
  year={2025}
}

@article{benerofe2025ai,
  title={AI Governance and the Verification Gap: A Framework for Law and Policy Under Computational Intractability},
  author={Benerofe, Steven},
  journal={Available at SSRN 5629290},
  year={2025}
}

@inproceedings{de2019does,
  title={Does object recognition work for everyone?},
  author={De Vries, Terrance and Misra, Ishan and Wang, Changhan and Van der Maaten, Laurens},
  booktitle={Proceedings of the IEEE/CVF conference on computer vision and pattern recognition workshops},
  pages={52--59},
  year={2019}
}

@inproceedings{NEURIPS2022_5474d9d4,
 author = {Gaviria Rojas, William and Diamos, Sudnya and Kini, Keertan and Kanter, David and Janapa Reddi, Vijay and Coleman, Cody},
 booktitle = {Advances in Neural Information Processing Systems},
 editor = {S. Koyejo and S. Mohamed and A. Agarwal and D. Belgrave and K. Cho and A. Oh},
 pages = {12979--12990},
 publisher = {Curran Associates, Inc.},
 title = {The Dollar Street Dataset: Images Representing the Geographic and Socioeconomic Diversity of the World},
 url = {https://proceedings.neurips.cc/paper_files/paper/2022/file/5474d9d43c0519aa176276ff2c1ca528-Paper-Datasets_and_Benchmarks.pdf},
 volume = {35},
 year = {2022}
}

@article{barker2018fifth,
  title={The fifth'CHiME'speech separation and recognition challenge: dataset, task and baselines},
  author={Barker, Jon and Watanabe, Shinji and Vincent, Emmanuel and Trmal, Jan},
  journal={arXiv preprint arXiv:1803.10609},
  year={2018}
}

@inproceedings{bender2021dangers,
author = {Bender, Emily M. and Gebru, Timnit and McMillan-Major, Angelina and Shmitchell, Shmargaret},
title = {On the Dangers of Stochastic Parrots: Can Language Models Be Too Big?},
year = {2021},
isbn = {9781450383097},
publisher = {Association for Computing Machinery},
address = {New York, NY, USA},
url = {https://doi.org/10.1145/3442188.3445922},
doi = {10.1145/3442188.3445922},
booktitle = {Proceedings of the 2021 ACM Conference on Fairness, Accountability, and Transparency},
pages = {610–623},
numpages = {14},
location = {Virtual Event, Canada},
series = {FAccT '21}
}

@inproceedings{al-ghanim-etal-2025-evaluating,
    title = "Evaluating the Robustness and Accuracy of Text Watermarking Under Real-World Cross-Lingual Manipulations",
    author = "Al Ghanim, Mansour  and
      Xue, Jiaqi  and
      Hastuti, Rochana Prih  and
      Zheng, Mengxin  and
      Solihin, Yan  and
      Lou, Qian",
    editor = "Christodoulopoulos, Christos  and
      Chakraborty, Tanmoy  and
      Rose, Carolyn  and
      Peng, Violet",
    booktitle = "Findings of the Association for Computational Linguistics: EMNLP 2025",
    month = nov,
    year = "2025",
    address = "Suzhou, China",
    publisher = "Association for Computational Linguistics",
    url = "https://aclanthology.org/2025.findings-emnlp.390/",
    doi = "10.18653/v1/2025.findings-emnlp.390",
    pages = "7396--7416",
    ISBN = "979-8-89176-335-7"
}

@InProceedings{pmlr-v81-buolamwini18a,
  title = 	 {Gender Shades: Intersectional Accuracy Disparities in Commercial Gender Classification},
  author = 	 {Buolamwini, Joy and Gebru, Timnit},
  booktitle = 	 {Proceedings of the 1st Conference on Fairness, Accountability and Transparency},
  pages = 	 {77--91},
  year = 	 {2018},
  editor = 	 {Friedler, Sorelle A. and Wilson, Christo},
  volume = 	 {81},
  series = 	 {Proceedings of Machine Learning Research},
  month = 	 {23--24 Feb},
  publisher =    {PMLR},
  url = 	 {https://proceedings.mlr.press/v81/buolamwini18a.html}
}

@techreport{nist_ai_rmf_2023,
  title        = {Artificial Intelligence Risk Management Framework ({AI RMF 1.0})},
  author       = {{National Institute of Standards and Technology}},
  institution  = {National Institute of Standards and Technology},
  year         = {2023},
  month        = jan,
  number       = {NIST AI 100-1},
  doi          = {10.6028/NIST.AI.100-1},
  url          = {https://nvlpubs.nist.gov/nistpubs/ai/nist.ai.100-1.pdf},
  note         = {U.S. Department of Commerce}
}

@article{li2020anti,
  title={Anti-compression JPEG steganography over repetitive compression networks},
  author={Li, Fengyong and Wu, Kui and Qin, Chuan and Lei, Jingsheng},
  journal={Signal Processing},
  volume={170},
  pages={107454},
  year={2020},
  publisher={Elsevier}
}

@article{wen2025sok,
  title={SoK: How robust is audio watermarking in generative AI models?},
  author={Wen, Yizhu and Innuganti, Ashwin and Ramos, Aaron Bien and Guo, Hanqing and Yan, Qiben},
  journal={arXiv preprint arXiv:2503.19176},
  year={2025}
}

@article{sorensen2024roadmap,
  title={A roadmap to pluralistic alignment},
  author={Sorensen, Taylor and Moore, Jared and Fisher, Jillian and Gordon, Mitchell and Mireshghallah, Niloofar and Rytting, Christopher Michael and Ye, Andre and Jiang, Liwei and Lu, Ximing and Dziri, Nouha and others},
  journal={arXiv preprint arXiv:2402.05070},
  year={2024}
}

@inproceedings{
kasirzadeh2024plurality,
title={Plurality of value pluralism and {AI} value alignment},
author={Atoosa Kasirzadeh},
booktitle={Pluralistic Alignment Workshop at NeurIPS 2024},
year={2024},
url={https://openreview.net/forum?id=AOokh1UYLH}
}

@inproceedings{yucer2022does,
  title={Does lossy image compression affect racial bias within face recognition?},
  author={Yucer, Seyma and Poyser, Matt and Al Moubayed, Noura and Breckon, Toby P},
  booktitle={2022 IEEE International Joint Conference on Biometrics (IJCB)},
  pages={1--10},
  year={2022},
  organization={IEEE}
}

\end{document}